\documentclass[]{spie}  
\usepackage[]{graphicx}

\title{Observing Strategies for the NICI Campaign to Directly Image Extrasolar Planets} 

\author{Beth Biller\supit{a} , \'Etienne Artigau\supit{b}, Zahed Wahhaj\supit{a}, Markus Hartung\supit{b}, Michael C. Liu\supit{a}, Laird M. Close\supit{c}, Mark R. Chun\supit{d}, Christ Ftaclas\supit{a}, Douglas W. Toomey\supit{e}, \& Thomas Hayward\supit{b}
\skiplinehalf
\supit{a}Institute for Astronomy, University of Hawaii, 2680 Woodlawn Drive, Honolulu, HI 96822 USA; \\
\supit{b}Gemini Observatory, Southern Operations Center, Association of Universities for Research in Astronomy, Inc., Casilla 603, La Serena, Chile; \\
\supit{c}Steward Observatory, University of Arizona, 933 North Cherry Avenue, Tucson, Arizona 85721, USA; \\
\supit{d}Institute for Astronomy, University of Hawaii, 640 North A'ohoku Place, 209 Hilo, Hawaii 96720-2700, USA; \\
\supit{e}Mauna Kea Infrared, LLC, 21 Pookela St., Hilo, Hawaii 96720, USA; \\
}

\authorinfo{Send correspondence to Beth Biller: bbiller@ifa.hawaii.edu}

\begin{document} 
  \maketitle 

\begin{abstract}
We discuss observing strategy for the Near
Infrared Coronagraphic Imager (NICI) on the 8-m Gemini South
telescope.  NICI combines a number of techniques to attenuate
starlight and suppress superspeckles: 1) coronagraphic imaging, 2)
dual channel imaging for Spectral Differential Imaging (SDI) and 3)
operation in a fixed Cassegrain rotator mode for Angular Differential
Imaging (ADI).  NICI will be used both in service mode and for a
dedicated 50 night planet search campaign. While all of these
techniques have been used individually in large planet-finding
surveys, this is the first time ADI and SDI will be used with a
coronagraph in a large survey.  Thus, novel observing strategies are necessary to conduct a viable planet search campaign.
\end{abstract}



\section{INTRODUCTION}
\label{sec:intro}  

To date, numerous adaptive optics surveys have attempted to directly image young extrasolar planets in the near infrared\cite{Biller07, Laf07a, Mas05}. While these surveys have placed strong statistical limits on extrasolar planet distributions, no planet has yet been imaged around a main sequence or pre-main sequence star.  Clearly, directly imaging planets is a nontrivial undertaking. It is important to weigh the survey instrument capability against theoretical predictions as to planet properties and distribution (e.g., from evolutionary models, radial velocity studies, etc.) and design a study which maximizes the chance of detecting a planet while also providing a scientifically interesting null result. In this paper, we will consider how best to maximize scientific output for a 50 night planet finding survey with the Near Infrared Coronagraphic Imager (NICI) at Gemini South.

NICI combines a number of techniques to attenuate
starlight and suppress superspeckles: 1) coronagraphic imaging, 2)
dual channel imaging for Spectral Differential Imaging (SDI)\cite{Mar05,Biller07} and 3)
operation in a fixed Cassegrain rotator mode for Angular Differential
Imaging (ADI) a.k.a. roll subtraction\cite{Liu04,Mar06}.  While coronagraphic imaging, SDI, and ADI have all been used individually in large planet-finding
surveys, this is the first time they will all be used together in a large survey. Thus, it is important to carefully consider how to best carry out an observing campaign with NICI. 

Both SDI and ADI techniques seek to separate real objects from speckles. SDI achieves this by exploiting a spectral feature in the desired target (previous implementations utilize the 1.6 $\mu$m methane absorption feature robustly observed in substellar objects cooler than 1400 K). Images are taken simultaneously both within and outside the chosen absorption feature. Due to the simultaneity of the observations, the star and the coherent speckle pattern are largely identical in both filters, while any faint companion with that absorption feature is bright in one filter and faint in the other. Subtracting the two images thus removes the starlight and speckle patterns while a real companion with the chosen absorption feature remains in the image. In other words, the off-absorption band image acts as an ideal reference point spread function (henceforth PSF) for the absorption band image.  Utilizing a signature spectral feature of substellar objects greatly reduces the number of false positives detected (e.g. a background object, while real, will drop out of the SDI
subtraction since it will not have methane absorption).

In previous surveys\cite{Biller07}, SDI was used at two discrete rotator angles for each object -- a real object on the sky will modulate with a change of rotator angle; however, speckles which are instrumental phenomena will not. Thus, by subtracting data taken at multiple roll angles, speckles are further attenuated and any real object in the frame will display a characteristic jump in roll angle.

ADI employs a similar strategy to build a high quality reference PSF to remove speckles. Instead of observing at discrete roll angles, ADI leaves the rotator off and allows the telescope optics to rotate on the sky (this works best at Cassegrain focus). In a sequence of images taken at different parallactic angles, a real companion will track on the sky with the parallactic angle, while speckles will move randomly. From a series of images, a reference PSF can be constructed for and subtracted from each individual image, attenuating quasi-static speckle structure. Combining both SDI and ADI techniques thus allows an even greater degree of speckle supression.

SDI is the base mode of the NICI instrument. Each exposure produces two frames, one from each of the two channels. Usually, each channel is operated with a different filter, one off the absorption feature (usually on the blue side), and one on the feature (usually redward of the first filter). Whether SDI is combined with ADI or used at discrete roll angles is the choice of the observer, and is not necessarily a straightforward choice. For more details regarding the NICI instrument itself, see Chun et al.\cite{Chun08}, this volume.

To develop the most appropriate observing strategy for the NICI science campaign, we must first
determine a number of quantities (some inherent to the objects observed, some properties of the 
NICI instrument itself), including:

1) the sky coverage and relative performance of the ADI plus SDI and SDI discrete roll angles observation modes (covered in Section 2)

2) the inherent luminosity of young planets and the required $\Delta$mag to image them (covered in Section 3.1)

3) the expected methane break strength for cool substellar objects and young planets, which will constrain how well our spectral difference works and how much of a $\Delta$mag boost
 we can get from it (also covered in Section 3.2) 

4) the achieved contrast of the NICI instrument as a function of separation from the primary star (relative $H$ band vs. $K$ band contrast discussed in Section 3.3, $H$ band contrast discussed in detail in Artigau et al.\cite{Art08} in this volume)

5) the achieved boost in contrast compared to the standard AO due to the spectral subtraction or due to azimuthal differential imaging (discussed in Artigau et al.\cite{Art08} in this volume)

6) the throughput of the NICI instrument at various wavelengths

In this document, we will therefore broadly discuss two major trades for the NICI campaign: operating with rotator off and on (ADI plus SDI vs. SDI discrete roll angles, respectively), and $H$ band vs. $K$ band performance.

\section{Rotator On vs. Rotator Off} 
\label{sec:rotonoff}

Previous direct detection surveys for planets have either utilized the multi-wavelength Spectral Differential Imaging technique (SDI) with one or more fixed rotator angles\cite{Biller07}
or utilized the Azimuthal Differential Imaging technique (ADI) at a single wavelength with the telescope rotator turned off\cite{Laf07a}. The NICI instrument is the first to combine both capabilities. Combining both techniques thus allows an even greater degree of speckle supression. However, while any observation can utilize the narrow-band methane filters to do SDI, ADI is not necessarily possible on all objects. To successfully perform ADI on an object, two conditions must be satisfied:

1) During the observing period, the target must rotate on the sky by at least 20$^{\circ}$ (corresponding to a rotation on the chip of 2 pixels at 1"). In ADI mode, a sequence of images is taken while the object moves through a variety of parallactic angles. For each image in the sequence, a reference PSF is built from the remaining images in the sequence. If the target has not rotated through a large enough angle, no images in the series will be appropriate for using as a reference PSF, because a companion object will not move sufficiently to prevent self-subtraction.

2) At the same time, the target must not rotate too much. For a given exposure time (e.g. 1 minute), if sky rotation is too much, an object on the sky at separation of greater than 2" will appear blurred in each image in the series (referred to hereafter as "shear"), also leading to lost companion flux.

Thus, only objects with enough total sky rotation ($>$20$^{\circ}$) but not too
high a rate of sky rotation (thus causing shear), are appropriate for rotator-off observations. Below, we discuss necessary criteria for rotator-off observations, the amount of sky accessible to rotator-off vs. rotator-on observations, and compare rotator-on vs. rotator-off performance.

\subsection{Analytical Considerations}
\label{sec:rotonoffanalytical}

\begin{figure}
	\begin{center}
  \begin{tabular}{cc}
	\includegraphics[width=3.2in]{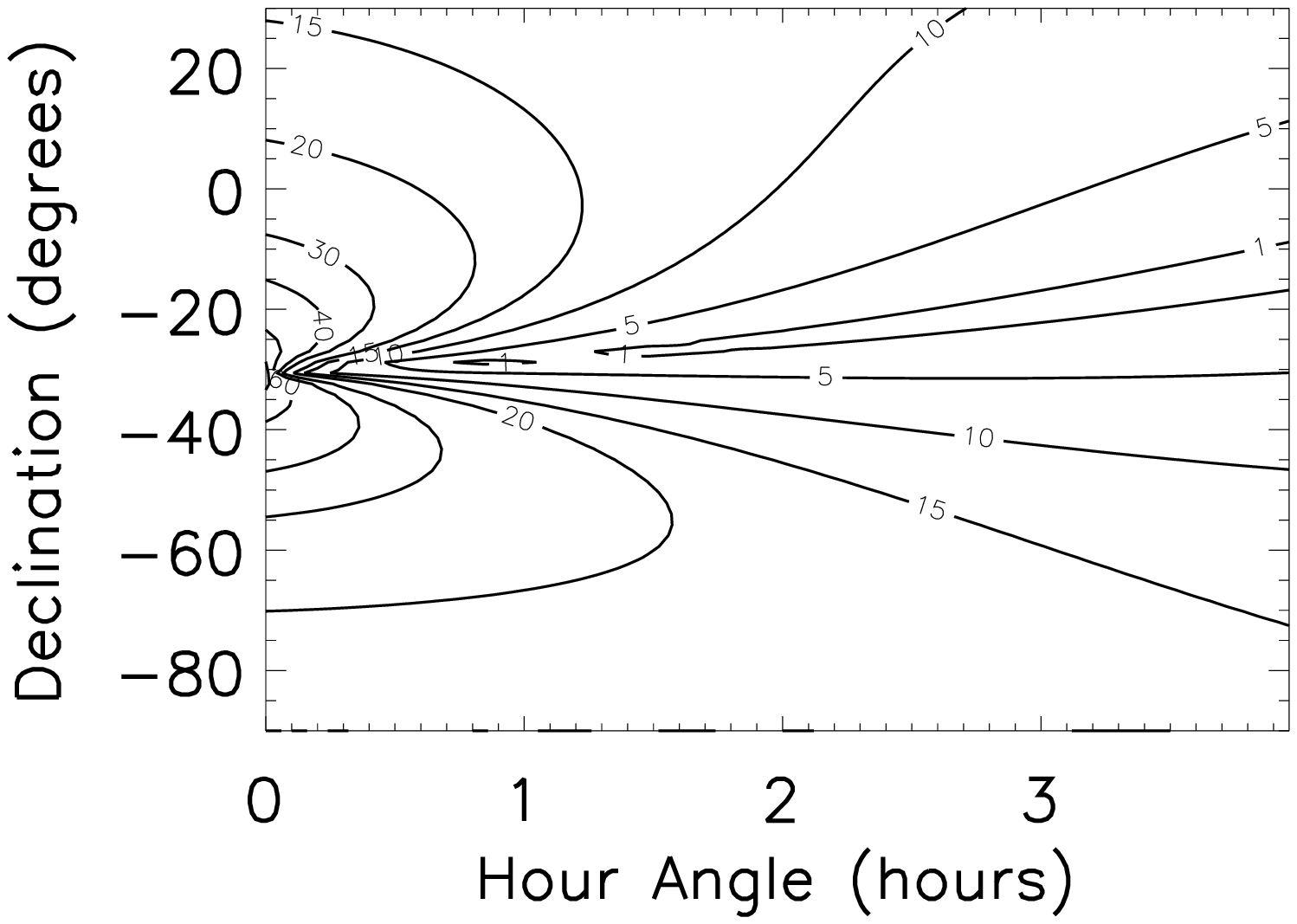} &
	\includegraphics[width=3.2in]{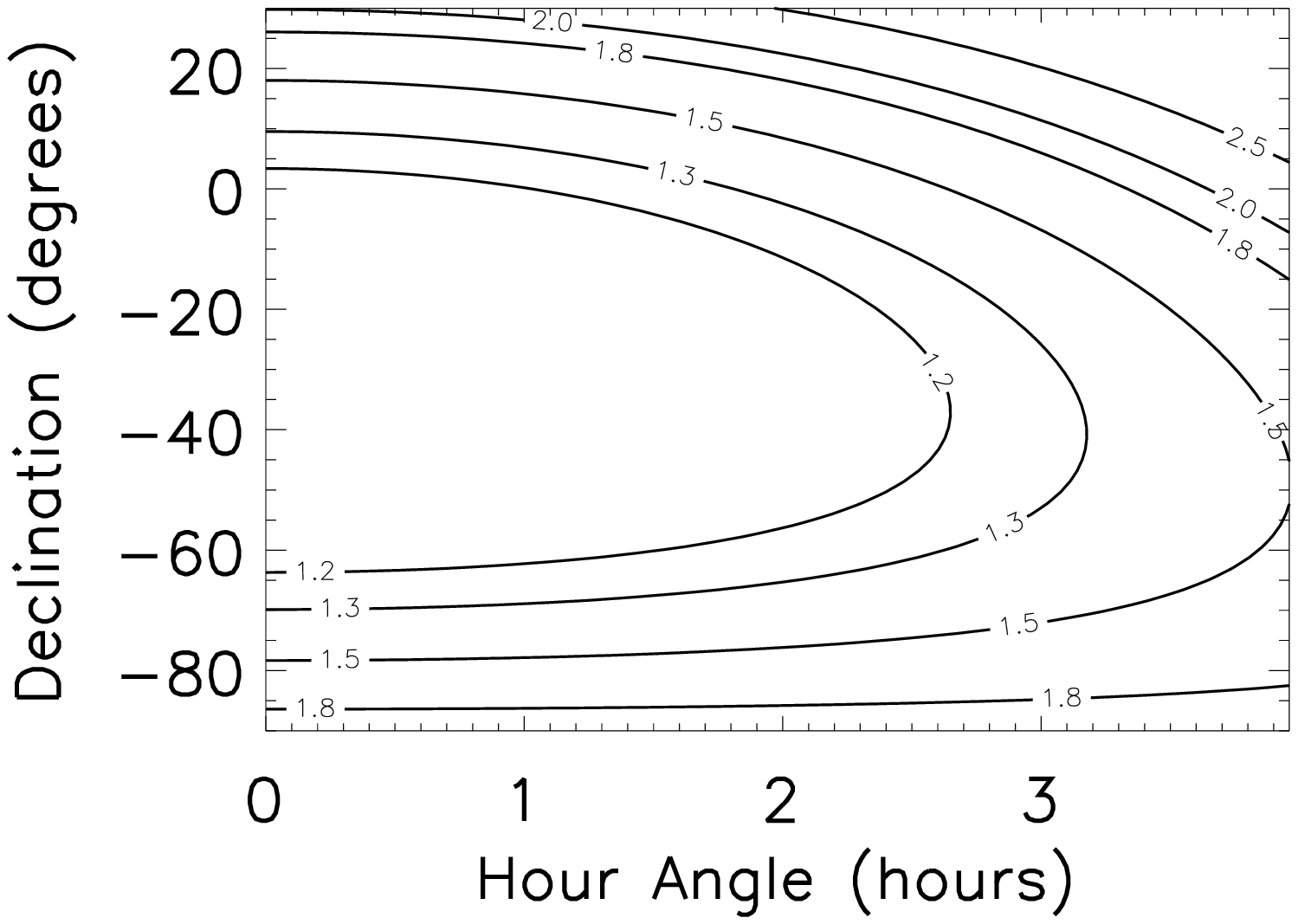} \\
	\end{tabular}
	\end{center}
	\caption{\label{fig:rotairmass}  Left: Rotation map on the sky (hour angle vs. declination) in an hour's observation at Gemini South. Hour angle plotted on the X axis of this figure is the hour angle at the beginning of the observation.  Contours are given at sky rotations of 60, 40, 30, 20, 15, 10, and 5$^{\circ}$. Targets with sky rotation of greater than 20$^{\circ}$ in an hour are observable with the ADI technique. Right: Airmass map on the sky for an hour's observation at Gemini South. Targets with airmass less than 2 are observable with the SDI technique at fixed rotator angles.
	}
\end{figure}

For any ADI observation, we need to balance having sufficient sky rotation to construct a reference PSF over a multi-image sequence vs. the acceptable shear per image. To constrain available sky rotation as a function of declination and hour angle, we plot maps on the sky (declination vs. hour angle) of rotation on the sky in an hour's observation in the left panel of Fig.~\ref{fig:rotairmass}. Hour angle plotted on the X axis of this figure is the hour angle at the beginning of the observation. (We have not plotted rotation over transit since parallactic angle changes so rapidly over transit that ADI observations are impossible due to shear -- see below.) For an object separated by 1" from the parent star, a sky rotation of 20$^{\circ}$ corresponds to a rotation on the chip of two pixels. With typical AO PSF FWHM$\sim$4 pix, this corresponds to about 1/2 FWHM of the AO PSF and is the minimum rotation necessary to ensure that a good reference PSF can be made. Thus, it is clear that only select portions of the sky are available to ADI at any given point in the night (although usable ADI time can be doubled by observing 
both before and after transit, omitting the period of highest sky rotation over transit). In contrast, most of the observable sky is available for SDI at discrete roll angles (see the right panel of Fig.~\ref{fig:rotairmass}, assuming the only constraint on whether SDI can be performed is having an airmass less than 2).

\begin{figure}
	\begin{center}
  \begin{tabular}{c}
	\includegraphics[width=3.2in]{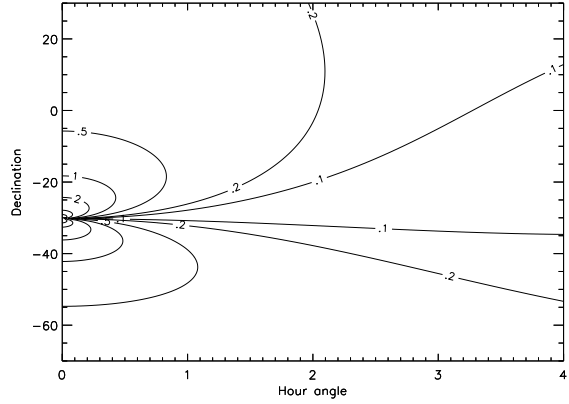} \\
	\end{tabular}
	\end{center}
	\caption{\label{fig:shear1}  Map of shear as a function of sky position (hour angle vs. declination) in a 1 minute exposure at Gemini South for an object 1" from the target. Contours are shown at shear levels of .1, .2, .5, 1, 2, and 5 pixels.
	}
\end{figure}	
	
Too much sky rotation can lead to blurring, or "shear", of a companion object image over one exposure. A map of shear at 1" from the primary in a 1 minute exposure is shown in Fig.~\ref{fig:shear1}. Thus, observing an object at -35 declination starting right after a transit will provide $>$30$^{\circ}$ of sky rotation, but also up to two pixels worth of shear for an object at 1" in a 1 minute exposure. Unfortunately, the portion of the sky that produces the greatest sky rotation will also produce the greatest shear, further reducing the part of the sky accessible with ADI.

It is unclear intuitively how much shear is too much. To estimate the efficiency loss due to shear, we used a simple geometric model. We assume a circular PSF, which will be elongated by shear.  The total energy in the PSF stays the same, but the peak flux decreases as PSF area increase.   
We model this elongated shape in a straightforward manner, assuming that the area of the PSF increases from a circular shape:

\begin{equation}
A_{noshear} = \pi r_{PSF}^2
\end{equation}

to the sheared final PSF (modeled as a rectangular sheared segment, with semi-circle edges):

\begin{equation}
A_{shear} = \pi r_{PSF}^2 + 2r_{PSF}*s
\end{equation}

where s is the total shear in pixels.
Thus, the peak flux F in the object is decreased by:

\begin{equation}
\frac{F_{shear}}{F_{noshear}} = \frac{A_{shear}}{A_{noshear}}
\end{equation}

and the sensitivity loss $\eta$ is given by:

\begin{equation}
\eta = 1 - \frac{F_{shear}}{F_{noshear}}
\end{equation}

\begin{figure}
	\begin{center}
  \begin{tabular}{c}
	\includegraphics[width=3.2in]{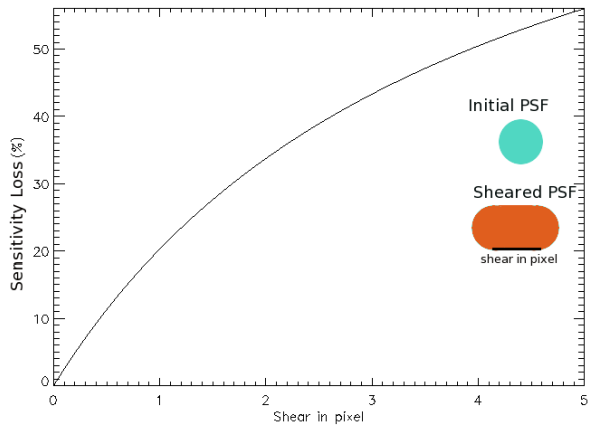} \\
	\end{tabular}
	\end{center}
	\caption{\label{fig:shear2}  Shear in pixel vs. sensitivity loss. Even up to 2 pixel shear leads to less than 50\% sensitivity loss.  
Initial and sheared PSF shapes are shown; the sheared PSF is modeled
as a rectangle (width given by the total shear in pixels) with two semicircular caps.
	}
\end{figure}

Shear in pixel vs. efficiency loss is presented in Fig.~\ref{fig:shear2}. Even for shears as much as 2 or 3 pixels, efficiency losses are less than 50\%, which makes highly sheared observations non-ideal, but still scientifically useful. One of the advantages of the NICI coronagraph compared to standard AO is the ability to have longer base exposure times because the coronagraph prevents saturation at the core of the target. In a situation where rotation would cause high shear, one can reduce base exposure time to reduce shear, but this negates the capability of the coronagraph to perform deep observations. Choosing a shorter base exposure time to reduce shear will also result in decreased depth of observation per image, an increase in telescope overhead, and additional read noise.
	
\subsection{On Sky Case Study}
\label{sec:rotonoffcasestudy}

Combining ADI with SDI theoretically allows for even more improvement in quasi-static speckle suppression compared to the SDI technique with discrete rotator angles. To quantify the relative performance of each technique, we compare results from two very similar datasets, one observed with ADI plus SDI, the other observed with SDI at discrete rotator angles.

On March 28, 2008, TWA7 was observed with NICI at Gemini South for 61 minutes prior to transit with the rotator off and 73 after transit with the rotator at two different discrete rotator angles (half at a rotator angle of 0$^{\circ}$, half at a rotator angle of 35$^{\circ}$). TWA7 (10 42 30.1 -33 40 16.2, J2000) is an M1 star, 8-10 Myr old, 55 $\pm$16 pc from Earth (assuming TW Hya membership \cite{ZS04}), with a background object at a separation of 3"\cite{Neu00} 
(2.5" separation at discovery), and 9.4 magnitudes fainter than the primary in $H$. A base exposure time of 30 seconds was used for each dataset. A 4\% filter centered on 1.58 $\mu$m was used in NICI's blue channel, and a 4\% filter centered on 1.65 $\mu$m was used in NICI's red channel. TWA7 rotated on the sky by 75$^{\circ}$ during the rotator off ADI plus SDI observation.  Thus, many images were available to construct an ideal reference PSF, but significant blurring was also observed in TWA7B (3" separation).

Each frame was flat-fielded and bad pixel corrected. The red frame from each red/blue frame image set was then scaled to the Airy pattern of the blue image, derotated, and aligned with its corresponding blue frame. A custom shift and subtract algorithm was used to align each red/blue frame image set on the speckle pattern. After precision alignment, the red frame was subtracted from the blue frame. For the SDI at discrete rotator angle dataset, all subtracted frames at each rotator angle were then median combined and the final 35 degree image was subtracted from the final 0 degree image.

\begin{figure}
	\begin{center}
  \begin{tabular}{c}
	\includegraphics[width=3.4in]{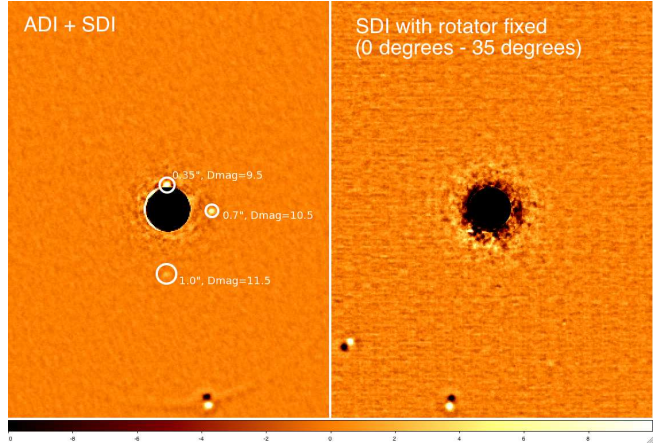} \\
	\end{tabular}
	\end{center}
	\caption{\label{fig:rotonoffimage}  Left: Final ADI plus SDI reduced image, Right: Final SDI with rotator fixed image. The inner .3" (under the coronagraphic mask) has been masked out in these images. Images are 4.7''$\times$6''.  The TWA7B background object is apparent at a separation of 3" in these images. Each frame was flat-fielded and bad pixel corrected. The red frame from each red/blue frame image set was then scaled to the Airy pattern of the blue image, derotated, and aligned with its corresponding blue frame. A custom shift and subtract algorithm was used to align each red/blue frame image set on the speckle pattern. After precision alignment, the red frame was subtracted from the blue frame. For the SDI at discrete rotator angle dataset, all subtracted frames at each rotator angle were then median combined and the final 35 degree image was subtracted from the final 0 degree image. For the ADI plus SDI dataset, a reference PSF was constructed for each subtracted image from the rest of the subtracted images in the sequence. The LOCI algorithm\cite{Laf07b,Art08} was used to construct the reference PSF. All frames which did not rotate on the sky by at least 1/2 FWHM at 1" were discarded when building the reference PSF. Simulated objects were also added into the ADI plus SDI datasets. Three simulated objects were added at separations of .35", .7", and 1", with $\Delta$mag = 9.5, 10.5, and 11.5 mag, respectively.  Note that the SDI with rotator fixed final image clearly has a higher noise floor than the ADI plus SDI final image; this is at least partly due to the fact that only one dither position was used with the SDI rotator fixed data.
	}
\end{figure}

For the ADI plus SDI dataset, a reference PSF was constructed for each subtracted image from the rest of the subtracted images in the sequence. The LOCI algorithm\cite{Laf07b,Art08} was used to construct the reference PSF. All frames which did not rotate on the sky by at least 1/2 FWHM at 1" were discarded when building the reference PSF. Final ADI plus SDI and SDI with fixed rotator angle images are presented in Fig.~\ref{fig:rotonoffimage}.

\begin{figure}
	\begin{center}
  \begin{tabular}{c}
	\includegraphics[width=3.4in]{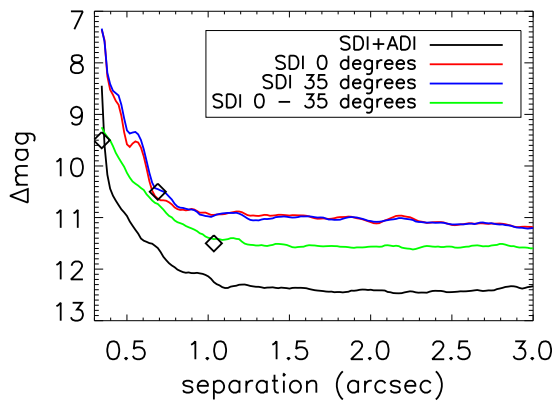} \\
	\end{tabular}
	\end{center}
	\caption{\label{fig:rotonoffcontrast}  $H$ band contrast curves for the ADI plus SDI and SDI at discrete rotator angles data. Contrast curves for each of these cases were generated by calculating the standard deviation within a 7 pixel wide annulus at each radial distance from the star. Magnitudes of contrast were scaled according to the unsaturated PSF of TWA7B in these images. For this dataset, ADI plus SDI achieves an extra magnitude of contrast compared to SDI at discrete rotator angles. This is somewhat due to the fact that this is a particularly optimal ADI observation within 3", and a somewhat sub-optimal SDI observation. The three simulated objects are overplotted here as diamond points.
	}
\end{figure}

Contrast curves for each of these cases were generated by calculating the standard deviation within a 7 pixel wide annulus at each radial distance from the star. Magnitudes of contrast were scaled according to the unsaturated PSF of TWA7B in these images. Contrast curves are presented in Fig.~\ref{fig:rotonoffcontrast}. For this dataset, ADI plus SDI achieves an extra magnitude of contrast compared to SDI at discrete rotator angles. This is somewhat due to the fact that this is a particularly optimal ADI observation within 3", and a somewhat sub-optimal SDI observation.

Simulated objects were also added into the ADI plus SDI datasets. Three simulated objects were added at separations of .35", .7", and 1", with $\Delta$mag = 9.5, 10.5, and 11.5 mag, respectively. These simulated objects are scaled images of TWA7B. A very strong methane break was assumed, thus, these objects appear only the blue filter. The objects were all detected at the 10-20 $\sigma$ level.

It is important to note that, while the ADI plus SDI dataset was well optimized, the SDI at discrete rotator angles dataset is somewhat sub-optimal. While the base exposure time for the ADI plus SDI dataset had to be limited to 30 seconds to reduce image shear, a longer base exposure time would have been optimal for the SDI at discrete rotator angle dataset. This would have allowed for deeper imaging, as well as a reduction in read noise and telescope overhead. In addition, the NICI detector suffers somewhat from checkerboard pattern noise, which is well removed in the ADI plus SDI dataset. However, as opposed to other implentations of SDI, only one dither position was used in the SDI discrete rotator angle dataset. This means that greater pattern noise remains after data reduction, resulting in a higher noise floor compared to the ADI plus SDI dataset. Finally, it is possible to get 75$^{\circ}$ of sky rotation only in a very small part of the sky; less sky rotation means less well constructed reference PSFs in the inner arcsec of the image. In other word, the inner arcsec in this ADI plus SDI dataset looks especially clean compared to more typical conditions. The noise floor encountered in the ADI plus SDI dataset at separations greater than 1" is likely due to read noise -- this dataset was taken with 4 nondestructive reads (NDR). Boosting the number of nondestructive will help bring down this limiting read noise floor.  Effort is currently being put into increasing the number of NDRs possible with NICI -- by the start of the NICI campaign, the NICI instrument will be able to be run with an arbitrary 
number of NDRs, thus significantly reducing the operational read noise floor.

\section{$H$ band vs. $K$ band performance}
\label{sec:HK}

Previous near-infrared direct detection searches for methane-rich planets have favored the $H$ band methane break over the $K$ band\cite{Biller07, Laf07a}. However, $K$ band also can provide a valuable search space, as well as follow-up capabilities. In this section, we investigate whether $H$ or $K$ provide more fertile ground for planet searches (e.g. inherent contrasts and methane break strengths in each band), as well as compare $H$ and $K$ band NICI performance. 

\subsection{$H$ vs. $K$ necessary contrasts}
\label{sec:HKcontrast}

We used recent young planet models\cite{For08} to constrain the inherent expected luminosity of young planets, and thus the contrasts necessary to detect them.  Planets around main sequence stars vs. planetary mass objects around very low mass stars and 
brown dwarfs likely have 
disparate formation mechanisms; the former likely form via core accretion or disk instability in a circumstellar disk, while the latter form (most likely) in a manner analogous to stars.  Thus, different initial
conditions must be taken into account when modeling each population.  Unfortunately, until 
recently, all available evolutionary models for planetary mass objects used "hot start" initial conditions --
starting from a very high initial entropy state.  This choice is likely appropriate for brown dwarfs, but 
not for true planets forming around a star (which require a lower entropy initial condition.)  Recently, Fortney et al. (accepted)\cite{For08} have modeled spectra for young planets using these more realistic 
initial conditions.   

\begin{figure}
	\begin{center}
  \begin{tabular}{cc}
	\includegraphics[width=3in]{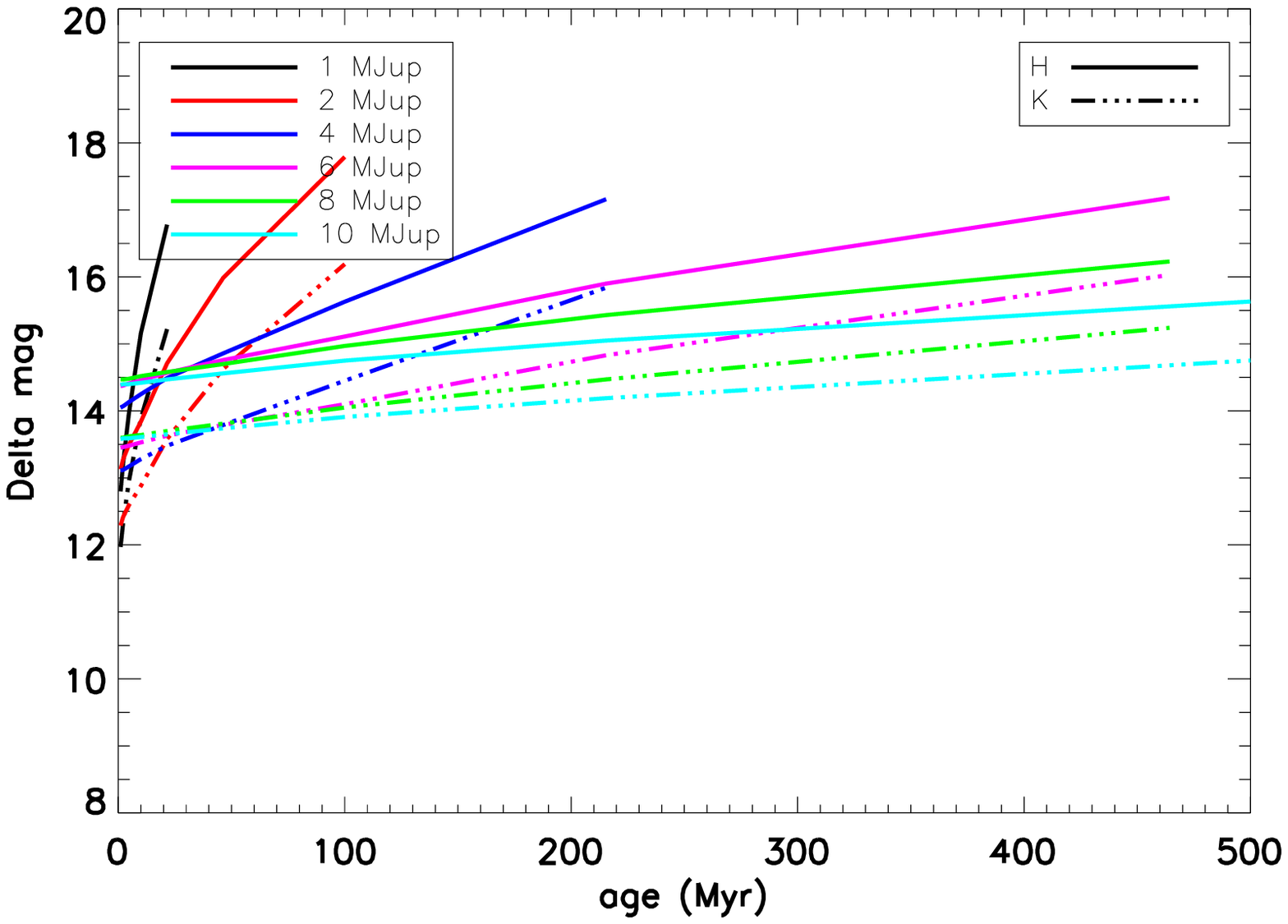} &
	\includegraphics[width=3in]{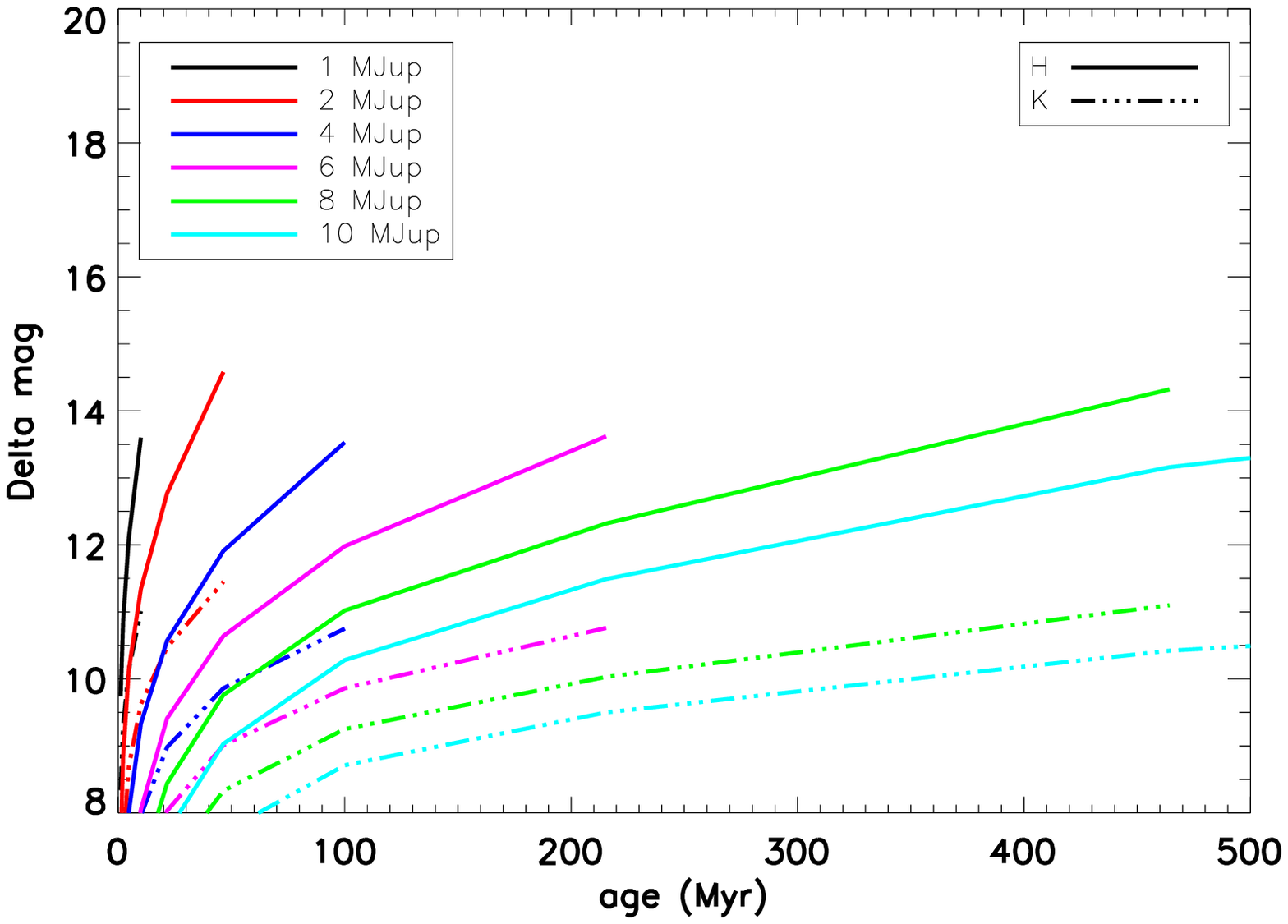} \\
	\end{tabular}
	\end{center}
	\caption{\label{fig:K2}  Required Contrasts to image planets of a range of masses for a K2 main sequence primary.  Left: Fortney et al. 2008 core accretion models.  Right: Fortney et al. 2008 hot start models.}
	\end{figure}

\begin{figure}
	\begin{center}
  \begin{tabular}{cc}
	\includegraphics[width=3in]{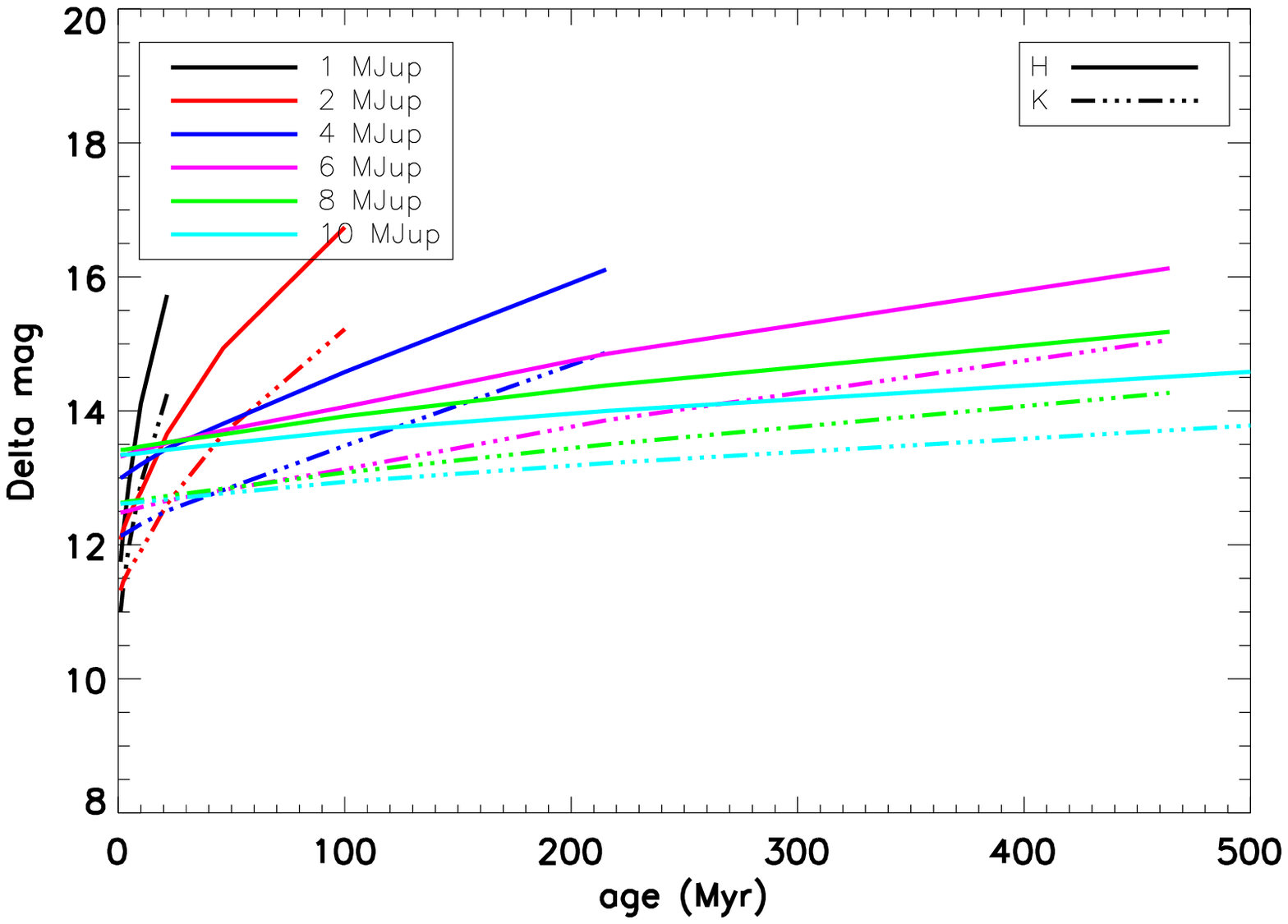} &
	\includegraphics[width=3in]{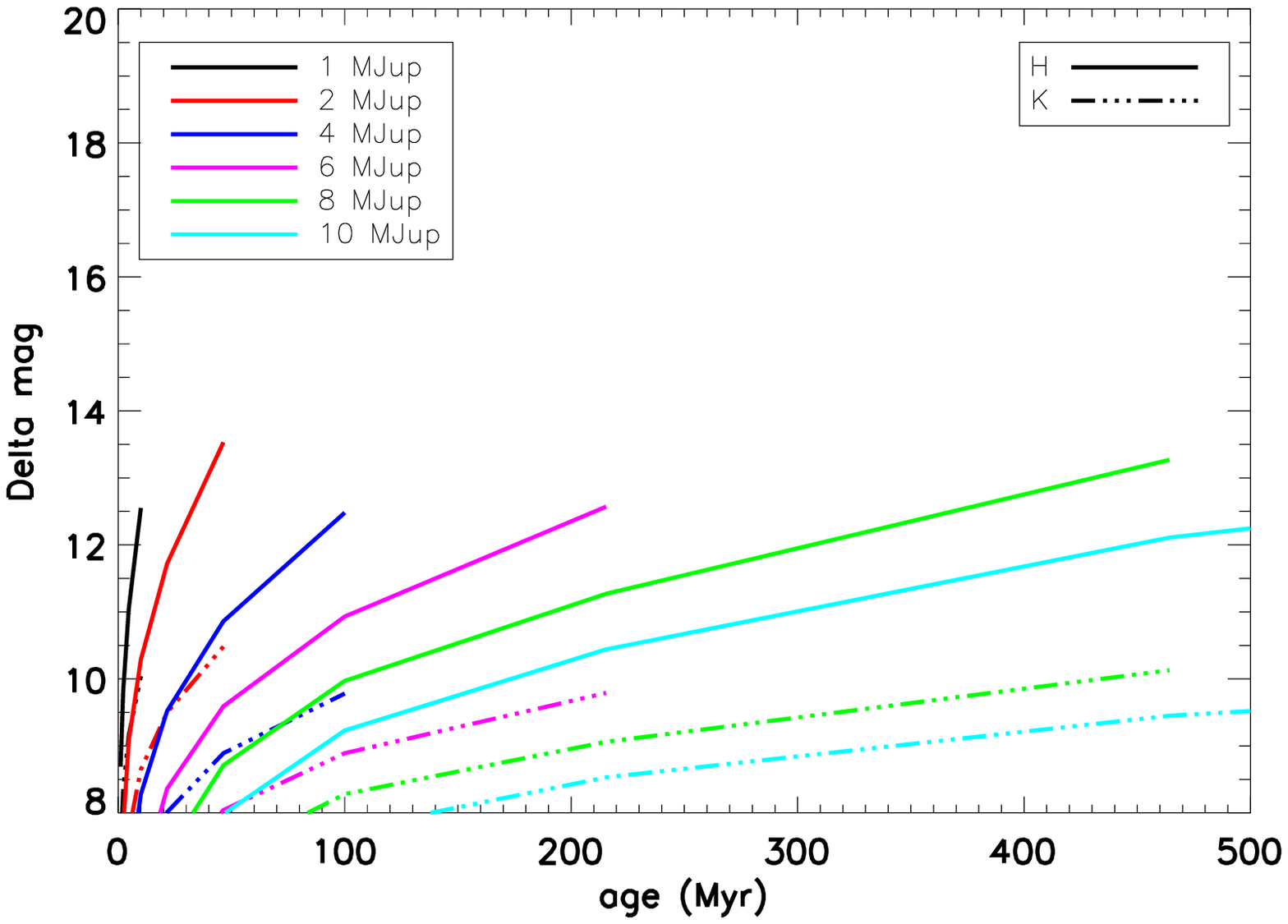} \\
	\end{tabular}
	\end{center}
	\caption{\label{fig:M0}  Required Contrasts to image planets of a range of masses for a M0 main sequence primary.  Left: Fortney et al. 2008 core accretion models.  Right: Fortney et al. 2008 hot start models.}
	\end{figure}

In order to evaluate $H$ vs. $K$ band performance in advance of the NICI science campaign, 
we calculated the necessary $\Delta$mag to image planets of varying masses in the $H$ and $K$ bands. Star-planet contrasts as a function of age for 1-10 M$_{Jup}$ planet models (core accretion on left, hot start on right, models from Fortney et al. accepted\cite{For08}) around main sequence K2 and M0 stars are plotted in Figs.~\ref{fig:K2} and ~\ref{fig:M0}.  (We used the simplifying assumption of main sequence primaries -- in reality, especially for the M0 primary, this assumption is probably not quite true.)  In general, much higher contrasts ($\Delta$mag=12-16 vs. $\Delta$mag=8-12) are required to 
image core accretion planets vs. hot start objects.  For objects of these ages and masses, $K$ band 
provides a more favorable $\Delta$mag than $H$.  In addition, these model planets are not especially overluminous in $K$ at very young ages. Previous surveys have operated on the assumption that young objects are especially bright compared to older ($>$300 Myr) objects. These new models suggest this brightness bump at early ages is considerably less than expected. Thus, upcoming direct planet surveys perhaps do not 
need to give very young objects ultra-high priorities.  Additionally, a viable strategy in light of these faint models would be to go very deep on fewer objects -- according to these predictions, a shallow survey is not likely to produce 
any detections or meaningful limits on planet distribution.

\subsection{$H$ vs. $K$ methane break strength}
\label{sec:HKbreak}

\begin{figure}
   \begin{center}
   \begin{tabular}{cc}
   \includegraphics[width=2in]{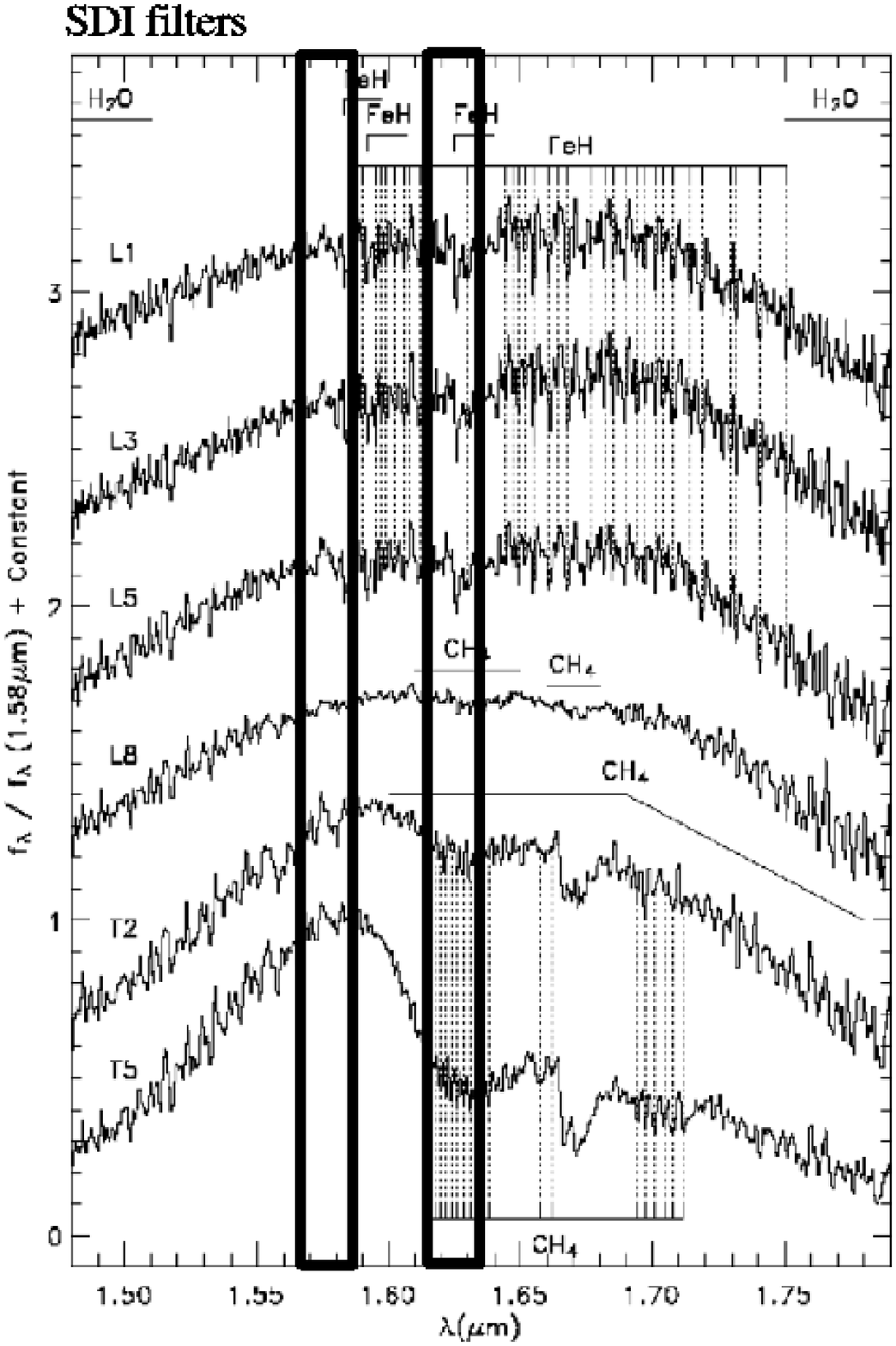} & 
   \includegraphics[width=2in]{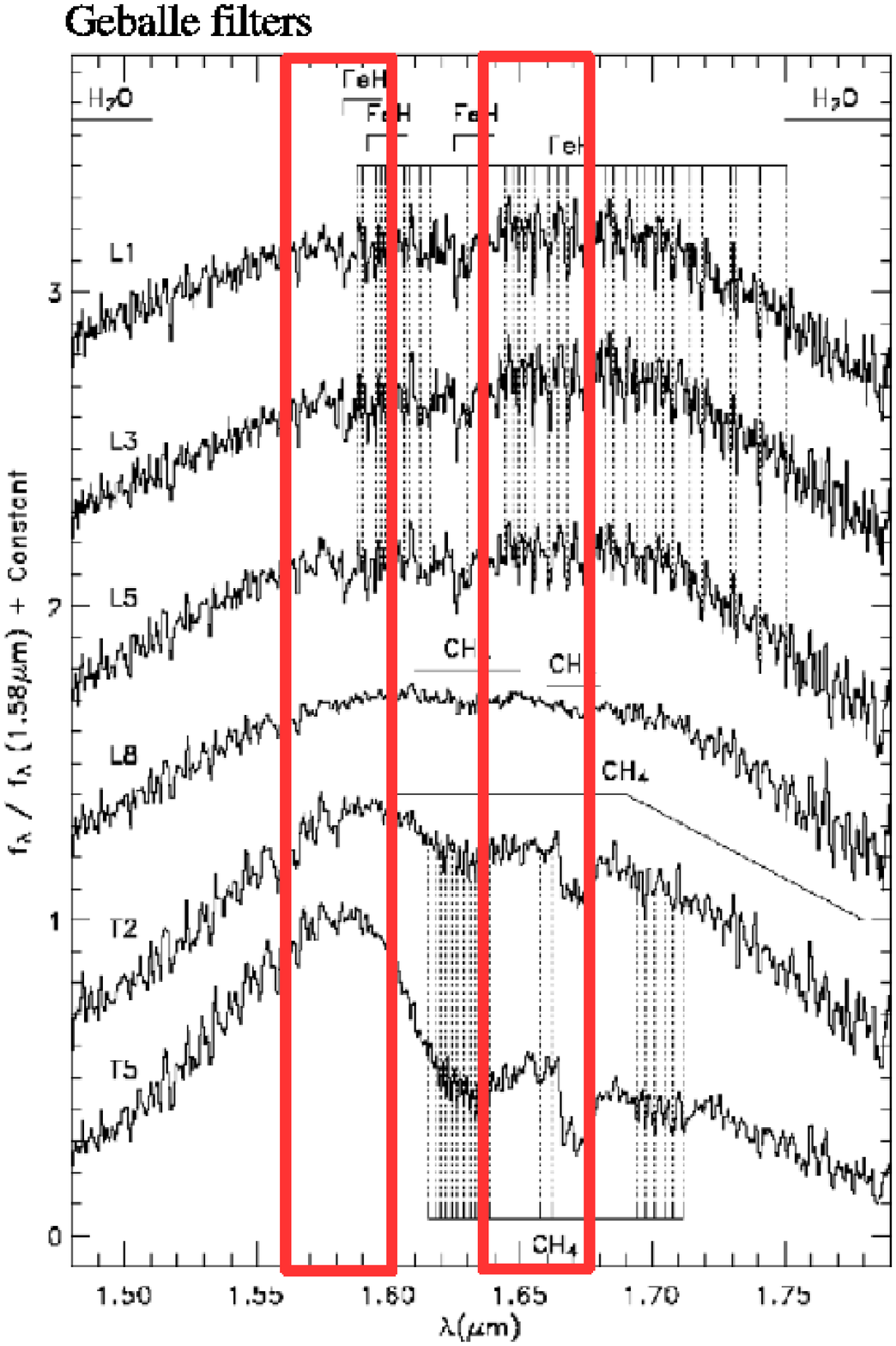} \\
   \includegraphics[width=2in]{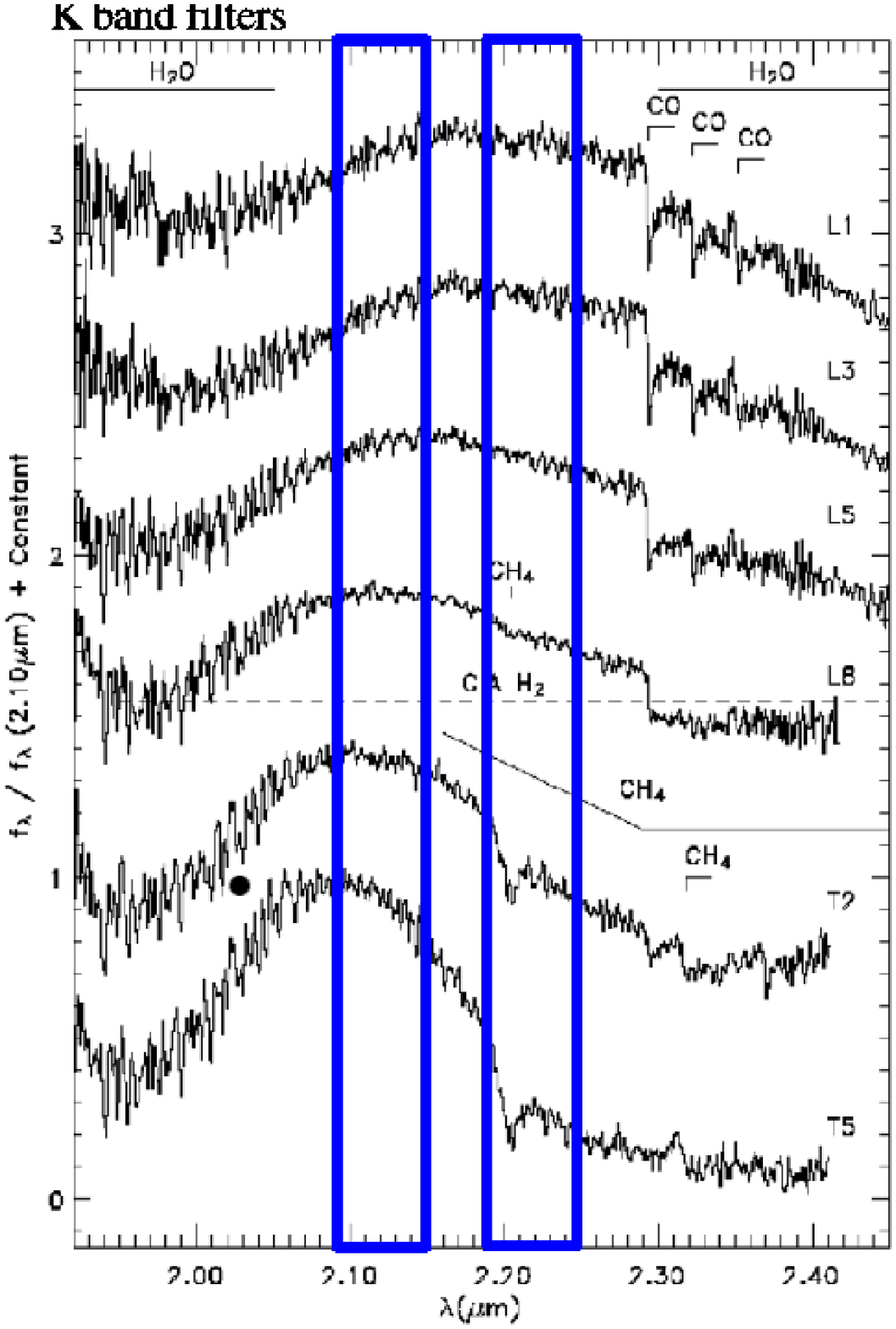} & \\
   \end{tabular}
   \end{center}
   \caption[] { \label{fig:spectra} $H$ band (top) and $K$ band (bottom)
spectra for a variety of L and T dwarfs.  Filter bandpasses are overlaid
for all three filter systems used to calculate spectral indices.
Methane absorption features
appear strongly in both the $H$ and $K$ band spectra.
Spectra from Cushing et al. 2005\cite{Cus05}.
}
   \end{figure}

Prominent methane absorption features appear both in the $H$ band 
(1.6 $\mu$m) and $K$ band (2.12 $\mu$m) spectra of T dwarfs 
(T$_{eff}$ $<$ 1200 K).  Representative 
L and T dwarf spectra\cite{Cus05} 
in both the $H$ and $K$ band are presented in Fig.~\ref{fig:spectra}.
While previous
methane imagers such as the SDI imager at the VLT and MMT have utilized
the $H$ band methane absorption feature, it is important to evaluate
the strength of both the $H$ and $K$ band features to determine which feature
best optimizes the performance of the NICI methane imager.

\begin{figure}
   \begin{center}
   \begin{tabular}{c}
   \includegraphics[width=3.2in]{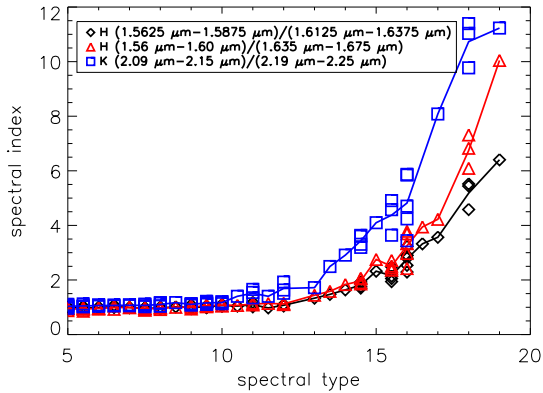}
   \end{tabular}
   \end{center}
   \caption[] { \label{fig:filters} Methane spectral indices for 2 $H$ band
and 1 $K$ band filter systems.  
We plot numerical spectral types on the x-axis; a numerical
type of 8 corresponds to a L8 spectral type, a numerical type of 16 
corresponds to a T6 spectral type, etc.  Spectral indices were calculated 
from spectra of 56 L dwarfs and 35 T dwarfs\cite{Kna04}.
Note that the $K$ band Methane break is 
stronger than the $H$ band Methane break for comparable filter systems.
}
   \end{figure}

To determine the relative strengths of the $H$ band and $K$ band methane 
absorption features we define a number of methane spectral indices of the form:

\begin{equation}
\frac{Flux_{unabsorbed}}{Flux_{absorbed}} = 
\frac{\int^{\lambda_2}_{\lambda_1} S_{\lambda} F({\lambda}) d\lambda}
{\int^{\lambda_4}_{\lambda_3} S_{\lambda} F({\lambda}) d\lambda}
\end{equation}

where S is the object flux and F is the filter transmission at wavelength 
$\lambda$.  Spectral indices were calculated from 
spectra of 56 L dwarfs and 35 T dwarfs\cite{Kna04}.
Spectra for these objects were obtained from 
Sandy Leggett's L and T dwarf archive\footnote[2]{http://www.jach.hawaii.edu/$\sim$skl/LTdata.html}.  Spectral index as a function of spectral type
is presented in Fig.~\ref{fig:filters}. (Although synthetic model spectra are available in Fortney et al. 2008, we did not calculate spectral indices for these model spectra, since methane opacities for the $H$ and $K$ bandheads are not well known. The dwarf L and T spectra used for this analysis are not exact analogues to young planets, but should have similar methane break strengths.) The standard deviation per spectral
type for these spectral indices range from $\sim$0.05 for late L's to 
$\sim$1.0 for mid to late T's.  In other words, the standard deviation of these
indices are fairly small and indices can be used to derive an accurate spectral
type for T dwarfs.

Two $H$ band indices were calculated.  The first of these,
$H$(1.5625 $\mu$m - 1.5875 $\mu$m)/(1.6125 $\mu$m-1.6375 $\mu$m), 
is the index calculated from the filters currently implemented in the 
VLT and MMT SDI imager. SDI filter transmission curves were convolved into these calculations.  The second $H$ band index, 
$H$(1.56 $\mu$m - 1.60 $\mu$m)/(1.635 $\mu$m-1.675 $\mu$m), is the methane
spectral index defined by Geballe et al. 2002\cite{Geb02}.  In this case, 
a tophat function was used for the filter transmission. One $K$ band index was also calculated -- K(2.09 $\mu$m - 2.15 $\mu$m)/
(2.19 $\mu$m - 2.25 $\mu$m). 
A tophat function was used for the filter transmission.  For mid to late T dwarfs, the $K$ band 
methane break is considerably stronger than the $H$ band break -- thus, future direct detection planet finding surveys should
strongly consider incorporating $K$ band methane break filters.

These indices only take into account the spectrum of a cool exoplanet 
companion.  In the real differential imaging case, we must also take into 
account the spectrum of the parent star.  In general, the star's spectrum 
will appear ``blue'' to any IR methane imager -- at these wavelengths, 
the IR starlight is from the Rayleigh-Jeans tail of its blackbody spectrum.  
This blue appearance will be slightly stronger in the 
K band compared to the $H$ band, but in general, the blue color of the star 
will be small for either band ($<$0.2 mag).

\subsection{$H$ band vs. $K$ band on sky performance}

\begin{figure}
   \begin{center}
   \begin{tabular}{c}
   \includegraphics[width=3.0in]{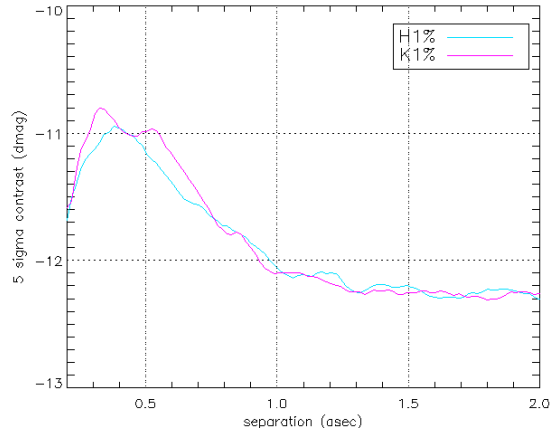}
   \end{tabular}
   \end{center}
   \caption[] { \label{fig:HKcontrast}  Contrast curves for $H$ and $K$ bands. Contrast curves for each of these cases were generated by calculating the standard deviation within a 7 pixel wide annulus at each radial distance from the star. Magnitudes of contrast were scaled according to the unsaturated PSFs of companion candidates\cite{Egg07}. Inside of 1", the $H$ band provides up to .4 mag better contrast than K, but otherwise contrast in these two bands appears to be similar.
}
   \end{figure}

To quantify the achieved contrast in $H$ vs. $K$ band, we compare results from two similar datasets, one taken in $H$ band, one in $K$. We observed HD 129642 for 15 minutes in the $K$ band (using Brackett $\gamma$ in one channel and $K$ continuum in the other) and 30 minutes in the $H$ band (using a 1\% filter centered on 1.58 $\mu$m in NICI's blue channel, and a 1\% filter centered on 1.65 $\mu$m in NICI's red channel). HD 129642 is a K2 star 28.6 pc distant surrounded by 5 companion candidates, one of which is likely bound\cite{Egg07}. A base exposure time of 30 seconds was used. The data were reduced according to procedures described in Section 2.2, and also in Artigau et al.\cite{Art08}, this volume.  Contrast curves were generated using the standard deviation in annular regions 7 pixels in diameter and scaling magnitudes according to the known companions in the field. $H$ and $K$ contrast curves are presented in Fig.~\ref{fig:HKcontrast}. Inside of 1", the $H$ band provides up to .4 mag better contrast than K (as expected because of the deeper exposure).  Otherwise contrast in these two bands appears to be roughly similar and limited by read noise beyond 1'' (8 NDRs were used with this dataset; boosting the number of NDRs will significantly decrease the read noise floor).  A more in depth comparison of $H$ vs. $K$ performance is planned for upcoming commissioning runs.

\section{Conclusions}
Careful scheduling is necessary to use ADI with SDI, since both shear per image and total rotation on the sky need to be taken into account. There will always be parts of the sky where ADI is completely unfeasible. In areas of the sky with enough target rotation but not too high a rate of rotation, ADI plus SDI does seem to provide considerable enhancements in speckle suppression over SDI at discrete rotation angles, although part of the difference for this particular comparison dataset may have resulted from the fact that the SDI dataset at discrete rotation angles was non-optimal. 

New, more realistic "cold start" models for young planets less than 300 Myr in age\cite{For08} suggest planets are fainter than originally predicted; thus, somewhat higher contrasts will be necessary to detect them (i.e., 12-16 mag vs. 8-12). Both theoretically necessary contrasts and predicted methane break strengths appear to be optimized in $K$ band relative to $H$ band. However, $H$ band contrast performance appears to be somewhat better than $K$ band inside 1". Additionally, it is important to consider total NICI instrument throughput in $H$ and $K$ bands when choosing between these, which we have not discussed in this document.

\acknowledgments     
Support for this work was provided by NASA through Hubble Fellowship grant HF-1204.01-A awarded by the Space Telescope Science Institute, which is operated by the Association of Universities for Research in Astronomy, Inc., for NASA, under contract NAS 5-26555.


\bibliography{report}   
\bibliographystyle{spiebib}   

\end{document}